\documentclass[11pt]{article}

\usepackage[preprint]{acl}

\usepackage{amsmath}
\usepackage{latexsym}
\usepackage{times}
\usepackage{xspace}
\setcitestyle{authoryear,round,semicolon}
\usepackage{array}
\usepackage{booktabs}
\usepackage{multirow}
\usepackage{algorithm}
\usepackage{algpseudocode}
\usepackage{amssymb}
\usepackage{float}
\usepackage[T1]{fontenc}
\usepackage[utf8]{inputenc}
\usepackage{microtype}
\usepackage{inconsolata}
\usepackage{graphicx}
\usepackage{xcolor}
\usepackage[normalem]{ulem}
\usepackage[colorinlistoftodos]{todonotes}

\newcommand{\cmark}{\ensuremath{\checkmark}}
\newcommand{\pmark}{\ensuremath{\circ}}
\newcommand{\xmark}{\ensuremath{\times}}
\newcommand{\model}{\textsc{CoWeaver}\xspace}

\newcommand\blfootnote[1]{%
  \begingroup
  \renewcommand\thefootnote{}\footnote{#1}%
  \addtocounter{footnote}{-1}%
  \endgroup
}
\newcommand\dagfootnote[1]{%
  \begingroup
  \renewcommand\thefootnote{\ensuremath{\dagger}}%
  \footnote{#1}%
  \addtocounter{footnote}{-1}%
  \endgroup
}

\title{CoWeaver: A Bi-directional, Learnable and Explainable Matching Engine for Mixed Human-Agent Science Collaboration}

\author{
  Jiayao Gu\textsuperscript{1,2,*} \quad
  Kexin Chu\textsuperscript{3,*} \quad
  Peidong Liu\textsuperscript{4} \\[0.4em]
  Yue Yang\textsuperscript{5} \quad
  Lynn Ai\textsuperscript{6} \quad
  Qi Zhang\textsuperscript{7} \quad
  Ling Yang\textsuperscript{8} \quad
  Tianyu Shi\textsuperscript{1}
}

\begin{document}
\maketitle
\blfootnote{%
    \textsuperscript{1}McGill University.
    \textsuperscript{2}Mila -- Qu\'ebec AI Institute.
    \textsuperscript{3}University of Connecticut.
    \textsuperscript{4}Sichuan University.
    \textsuperscript{5}Stanford University.
    \textsuperscript{6}Gradient.
    \textsuperscript{7}Independent.
    \textsuperscript{8}Princeton University.
    \textsuperscript{*}These authors contributed equally to this work.
    Correspondence to: \texttt{tianyu.shi3@mcgill.ca }.
}

\begin{abstract}
LLM-based agents excel at writing articles, coding and information retrieval. However, they fail to form strong collaborations within the scientific community due to the bidirectional, dynamic nature of the problem and a high demand of decision interpretability. We proposed \model, a bidirectional, learnable and explainable algorithm to match scientists and form strong collaborations within a human-agent network. \model matches candidates and requesters through filling capability gaps and filters candidates through a two-stage ranking step. Finally, the model explores newcomers by maintaining uncertainty-aware capability estimates and updating them through requester's feedback. We show that the selection mechanism of combining both exploration (UCB) and greedy of \model exceeds the greedy-only mechanism - the analytical best solution - on 6 out of the 20 tasks and performed on par with the greedy-only mechanism in terms of selecting the best candidate. We compared \model baselines in terms of matching quality and efficiency. \model outperforms baselines on all metrics. \dagfootnote{Code and data:
\url{https://anonymous.4open.science/r/CoWeaver-8B0B}.}
\end{abstract}

\section{Introduction}

Agents developed on top of Large language models (LLMs) have become increasingly capable at answering questions, writing code, and retrieving information. However, these systems still struggle with collaboration-formation tasks in social and scientific settings where sustained coordination, incentive alignment, and feasibility checks are required.
Recent evaluations of LLM collaboration agents show that, although promising, these systems often remain unstable under realistic coordination pressure and can still fail to form durable, high-value human-agent partnerships \citep{huang-etal-2025-enable,sun-etal-2025-collab,zhang2024exploring,jhamtani-etal-2025-llm}.
This paper studies \textbf{collaboration matching} in mixed human-agent networks, with scientific collaboration as a motivating setting.

The need for such matching is especially visible for early-career researchers, who often know the research bottleneck before they know the right collaborator. In a HuMAS academic collaboration-agent survey, 44\% of respondents reported that they would first seek help from mentors or domain experts when facing a bottleneck, and 81\% described their most important collaborator as partially or highly complementary in expertise. These responses mirror the technical requirements below: collaboration matching should identify complementary capabilities, reason about value to both sides, and screen feasibility rather than rank globally similar profiles. Additional motivation evidence is provided in Appendix~\ref{app:user-survey}.

Scientific work is rarely a single-agent problem.
A project may require domain expertise, experimental infrastructure, data access, writing experience, and sustained execution, while any individual researcher or agent may cover only part of this space.
Existing systems only partially address this challenge.
Recommender systems and multi-stakeholder recommendation model preferences, exposure, cold start, and platform-side objectives \citep{adomavicius2005toward,bobadilla2013recommender,abdollahpouri2020multistakeholder,burke2025decentering}; expert team formation emphasizes skill coverage and collaboration structure \citep{lappas2009team,saeedi2025survey}; stable and learned matching markets reason about two-sided preferences under stability or uncertainty assumptions \citep{gale1962college,li2025bandit}; and recent LLM-agent systems study agent coordination and human-model cooperation \citep{huang-etal-2025-enable,jhamtani-etal-2025-llm,wang-etal-2025-anymac,sun-etal-2025-collab}.
Collectively, this body of work only partially covers the requirements for real collaboration formation: existing methods typically specialize in one or two dimensions of capability coverage, both-side value, hard feasibility constraints, or cold-start handling.
What is missing is an integrated model that handles all four together for mixed human--agent matching: filling concrete capability gaps, modeling value on both sides of a match, enforcing non-negotiable feasibility constraints, and supporting under-observed participants.
This motivates the need for a dedicated matching model.

The central difficulty is that collaboration formation is not simply a ranking problem.
A suitable candidate must be able to fill the requester's missing capabilities, but the task must also provide enough value for the candidate.
Moreover, a match between the requester and the candidate changes the future state of many factors -- communication cost, coordination friction, likelihood to complete the task, trust between collaborators, subsequent feedback and more.
This makes the problem bidirectional, dynamic, and unpredictable.
It also makes model's interpretability an important factor -- users are unlikely to accept collaboration suggestions unless the system can state what gap is being filled, what incentive is being satisfied, and which constraints have been checked.

Our design is motivated by a simple principle: a method that can actively create strong connections between potential collaborators for any specific task. Within a scientific community, this means jointly considering human researchers, proxy agents that represent user goals, service agents that provide specialized capabilities, and reusable skills that can be invoked during work.
The matching engine should therefore behave less like a static recommender and more like a coordination layer that observes the network, identifies complementary opportunities, filters unsafe or infeasible links, and learns from collaboration outcomes.

We introduce \model, a bidirectional, learnable and explainable algorithm to match requester and candidates to form strong scientific collaborations in a mixed scientists-agent network.
The system represents humans and agents in a shared semantic state space, where each actor is represented by a compact vector of capabilities, task intents, constraints, and collaboration history, estimates bidirectional match value with \textsc{MapScore}, and separates hard constraints from utility ranking through explicit gate variables.
It then uses a two-stage pipeline: an analytical coarse stage ranks candidates efficiently by capability complementarity and motivation fit, while a simulation-based re-ranking stage further filters candidates according to soft traits including but not limited to time-zone difference, scheduling difference, collaboration style.
Finally, the model maintains uncertainty-aware capability estimates and updates them from feedback, allowing cold-start candidates to be explored while gradually converging toward reliable matches.

Our main contributions are:

\begin{itemize}
  \item We propose a bidirectional collaboration-matching mechanism for mixed human-agent networks, where capability gaps, participant motivation, uncertainty, and hard feasibility constraints are modeled jointly.
  \item We design \textsc{MapScore} and a two-stage planning pipeline that combines interpretable analytical matching with simulation-based collaboration assessment, aiming to balance scalability and process-level realism.
  \item We implement \model, an end-to-end matching engine for human-agent mixed network in science collaboration.
\end{itemize}

\section{Method}
\label{sec:method}

\begin{figure*}[t]
    \centering
    \includegraphics[width=\linewidth]{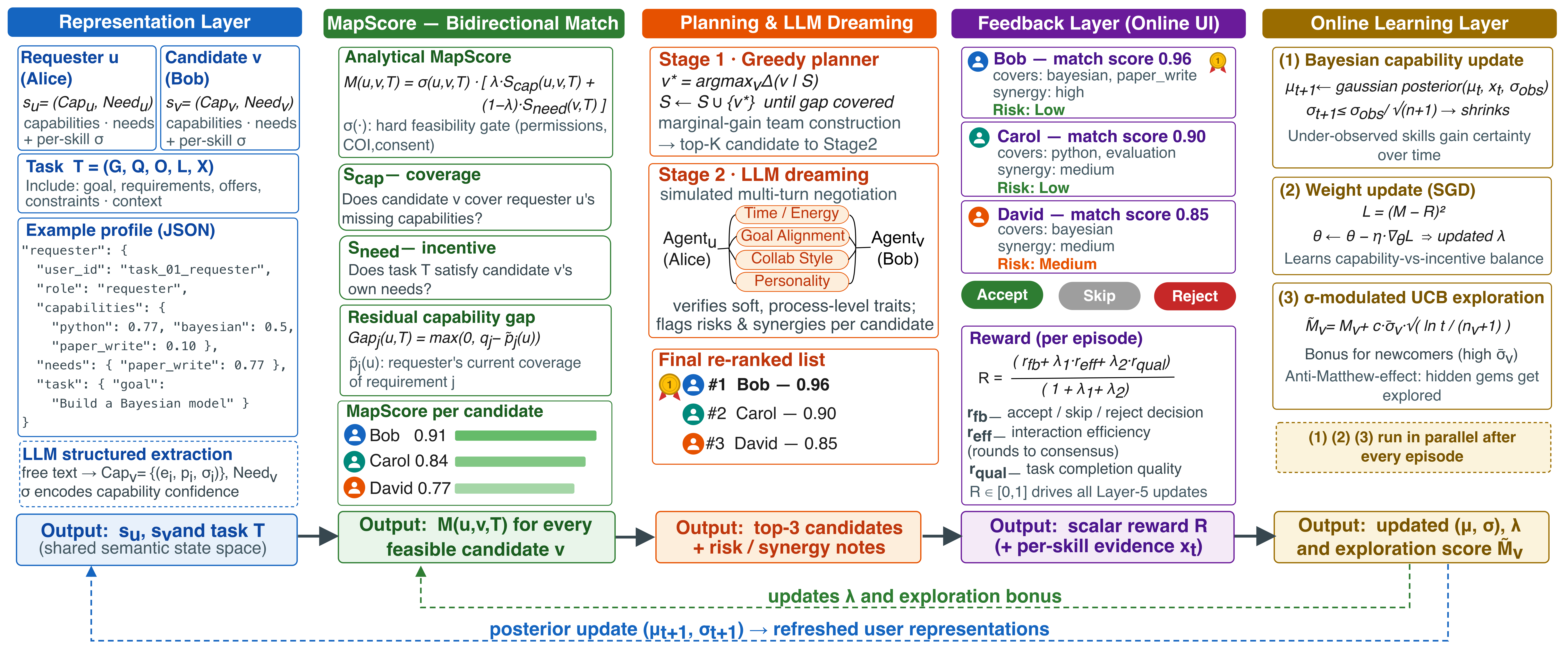}
    \caption{Overview of \model as an bidirectional, learnable and explainable matching engine for mixed human-agent scientific collaboration.}
    \label{fig:model_framework}
\end{figure*}

\subsection{Overview}

Given a requester $u$, a task $T$, and a candidate pool $\mathcal C$, \model{} returns either a ranked list of collaborators or a small team proposal.We refer to each selectable human, proxy agent, service agent, or reusable skill as a \emph{candidate}.
Unlike one-shot recommendation, the system is organized as a closed-loop matching pipeline: it represents heterogeneous candidates, estimates bidirectional collaboration value, decomposes scores into interpretable factors, plans candidate connections, refines promising matches through \textsc{LLM Dreaming}, and updates its estimates from feedback.

Figure~\ref{fig:model_framework} provides the design overview. We subsequently introduce each layer of our design from \S 2.2 to 2.5.
The key idea is to separate fast analytical matching from slower process-level reasoning.
MapScore serves as an analytical matching score for estimating whether a candidate is useful to the requester and whether the task is valuable to the candidate.
The closed-form attribution layer exposes which requester-side requirements and candidate-side needs drive the score.
Planning uses this score to construct a shortlist or team proposal, while \textsc{LLM Dreaming} is reserved for top candidates where communication risks, role conflicts, and priority mismatch are more consequential.
The feedback loop then updates capability estimates, uncertainty, and scoring weights.

\subsection{Representation Layer}

Each user $v$ is represented by capabilities, needs, and constraints:
\[
\mathbf s_v=(\mathbf{Cap}_v,\mathbf{Need}_v).
\]
The capability set is open-ended:
\[
\mathbf{Cap}_v=\{(\mathbf e_i^v,p_i^v,\sigma_i^v)\}_{i=1}^{n_v},
\]
where $\mathbf e_i^v$ is a semantic embedding, $p_i^v\in[0,1]$ is an estimated proficiency, and $\sigma_i^v$ is uncertainty.
The needs records how much user $v$ needs assistance in domain $k$:
\[
    \mathbf{Need}_v = \{n_v^k\}_{k=1}^{k=K} \in [0,1]^K
\]
This shared representation allows humans, proxy agents, service agents, and reusable skills to enter the same matching space while retaining type-specific constraints. A task is represented as
\[
T=(G_T,\mathbf Q_T,\mathbf O_T,\mathbf L_T,\mathbf X_T),
\]
where $G_T$ is the goal, $\mathbf Q_T$ contains capability requirements, $\mathbf O_T$ contains offers or incentives, $\mathbf L_T$ contains hard constraints, and $\mathbf X_T$ provides context.
The distinction between requirements and offers is central: requirements support requester-side gap estimation, while offers support candidate-side participation value.
Additional representation details are given in Appendix~\ref{app:semantic-matching}.

\subsection{MapScore: Bidirectional Matching}

MapScore is the core analytical matching score. It first enforces non-compensatory feasibility and then ranks feasible pairs by bidirectional collaboration value.
It is not a similarity score; it estimates task-conditioned collaboration value by combining two asymmetric questions: whether candidate covers the requester's $\mathbf{Need}_u$, and whether task $T$ satisfies the candidate's $\mathbf{Need}_v$.
Formally,
\[
\begin{aligned}
M(u,v,T)=
&\ \sigma(u,v,T)\Big[
\lambda S_{\mathrm{cap}}(u,v,T) \\
&\quad +(1-\lambda)S_{\mathrm{need}}(v,T)
\Big],
\end{aligned}
\]
where $S_{\mathrm{cap}}$ measures capability-gap coverage, $S_{\mathrm{need}}$ measures participation-value fit, $\lambda\in(0,1)$ is learned from feedback, and $\sigma(u,v,T)$ is a hard feasibility gate.

For a task requirement $j$, we first estimate the requester's current coverage $\tilde p_u^j$ through semantic matching and define the residual gap as
\[
\mathrm{Gap}_j(u,T)=\max(0,q_j-\tilde p_u^j).
\]
The capability term $S_{\mathrm{cap}}$ then aggregates how much of this residual gap candidate $v$ can cover.
The participation term $S_{\mathrm{need}}$ matches task offers against candidate needs, preventing the system from treating candidates as passive resources.
The full attention-based computation of both terms is provided in Appendix~\ref{app:semantic-matching}. Because both soft utility terms aggregate explicit requirement and need factors, MapScore also supports the closed-form attribution which is further described in Appendix \ref{app:mapscore_closed_form}. 


\subsection{Planning and LLM Dreaming}
For one-to-one matching, the planner computes MapScore for each candidate, filters infeasible pairs, and sorts the remaining candidates by score.
For one-to-$N$ matching, the planner maintains a residual capability gap $\mathbf R$ and greedily adds the feasible candidate with the largest marginal value:
\[
\begin{aligned}
\Delta(v\mid \mathcal S)
=&\ \sigma(u,v,T)\Bigg[
\rho
\frac{\sum_j \min(R_j,c_v^j)}
{\sum_j R_j+\epsilon} \\
&\quad +(1-\rho)S_{\mathrm{need}}(v,T)
\Bigg],
\end{aligned}
\]
where $\rho$ balances residual capability coverage and candidate-side value (Algorithm \ref{algo:greedy}).

\begin{algorithm}[h]
\caption{Residual-gap team construction}
\label{algo:greedy}
\begin{algorithmic}[1]
\State $\mathcal S\leftarrow\varnothing$; $\mathbf R\leftarrow\mathbf{Gap}(u,T)$
\While{$|\mathcal S|<n_{\max}$ and $\|\mathbf R\|_1>\epsilon$}
  \State $v^*\leftarrow\arg\max_{v\in\mathcal C\setminus\mathcal S}\Delta(v\mid\mathcal S)$
  \If{$\Delta(v^*\mid\mathcal S)\le 0$}
    \State break
  \EndIf
  \State $\mathcal S\leftarrow\mathcal S\cup\{v^*\}$
  \State $R_j\leftarrow\max(0,R_j-c_{v^*}^j)$ for all $j$
\EndWhile
\State \Return $\mathcal S$
\end{algorithmic}
\end{algorithm}

This residual-gap strategy gives a scalable approximation to coverage-based team formation \citep{lappas2009team}.

Analytical scores are efficient, but collaboration can still fail due to soft process-level factors.
We therefore apply \textsc{LLM Dreaming}. For each requester–candidate pair, we instantiate two role-conditioned agents using their structured profiles: capabilities, needs, availability, timezone, deadline pressure, collaboration style, communication preference, personality notes, and ranked priorities. The agents conduct a short simulated negotiation about the target task, including concrete work-allocation tradeoffs over a 100-point effort budget, so that the model can expose latent frictions not captured by the analytical score alone. The dreamed interaction is then judged along four soft-compatibility dimensions: time/energy feasibility $\gamma_v$-- whether the candidate’s stated availability supports the task workload, priority or goal alignment $\alpha_v$--whether the requester and candidate agree on what matters for the collaboration, collaboration-style $\kappa_v$ (e.g., async vs. synchronous work, structured vs. flexible execution), and personality fit $\phi_v$--whether their negotiation and communication styles are mutually productive rather than one-sided or conflict-prone. The judge returns a simulated score $S_\mathrm{sim}(u,v,T)$ together with a risk, and synergy recommendation (Fig. \ref{fig:model_framework}):
\[
S_\mathrm{sim}=
0.50\,\phi_v +
0.20\,\kappa_v +
0.20\,\alpha_v +
0.10\,\gamma_v ,
\]
The final ranking is
\[
R(a)=\eta M_a+(1-\eta)S_{\mathrm{sim}}(u,v,T),
\quad \eta\in[0,1],
\]
where $M_a$ is the analytical score of the candidate or team.
The output includes the ranked suggestion together with covered gaps, satisfied needs, passed constraints, and simulation-identified risks.

\subsection{Feedback Learning and Exploration}

The output of the LLM dreaming stage is presented to the requester as a ranked set of candidate recommendation cards. For each candidate, the interface displays the final score, the analytical matching components --capability-gap coverage and need satisfaction, the LLM-derived synergy and risk labels. The requester then provides feedback through: accepting, skipping, or rejecting a candidate. These actions are mapped into a structured reward signal used by the online-learning module: The final scalar reward is computed as
\[
R =
\frac{
r_{\mathrm{fb}}
+ \lambda_1 r_{\mathrm{eff}}
+ \lambda_2 r_{\mathrm{qual}}
}{
1 + \lambda_1 + \lambda_2
},
\]
where \(r_{\mathrm{fb}}\) is the immediate user reaction,
\(r_{\mathrm{eff}}\) measures interaction efficiency, and
\(r_{\mathrm{qual}}\) captures completion- and rating-based quality.
This reward is then used to update the candidate's Bayesian capability
estimates and selection counts, so subsequent recommendations reflect the
requester's observed preferences.
This feedback updates three parts of the matcher.
First, capability-specific evidence updates proficiency and uncertainty estimates, so under-observed capabilities become more certain as they are validated.
Second, the balance parameter $\lambda$ is updated to reflect whether capability coverage or participation value better predicts success in the current environment.
Third, candidate retrieval uses uncertainty-aware exploration to reduce cold-start lockout.

Concretely, retrieval can use an upper-confidence selection heuristic \citep{auer2002finite}:
\[
\mathrm{Select}_t(v)=
M(u,v,T)+c\,\bar\sigma_v
\sqrt{\frac{\log(t+1)}{n_v+1}},
\]
where $\bar\sigma_v$ is the candidate's mean uncertainty and $n_v$ is the number of previous interactions.
This heuristic gives new or under-observed humans, agents, and skills a chance to enter the shortlist, while posterior updates gradually shift the system from exploration toward exploitation.
Detailed update equations are provided in Appendix~\ref{app:feedback-updates}.

\section{Evaluation}
\label{sec:eval}

We evaluate \model{} as a task-conditioned collaboration matcher along four questions.
First, we test whether analytical online selection can recover high-quality collaborators under a limited interaction budget.
Second, we examine whether \textsc{LLM Dreaming} improves process-level feasibility beyond capability matching.
Third, we test whether MapScore explanations are faithful to the analytical matching decision.
Finally, we compare against a negotiation-style agentic baseline to assess end-to-end matching quality and efficiency.

\paragraph{Research questions.}
RQ1 asks whether analytical online selection can recover high-quality collaborators under limited rounds.
RQ2 asks whether \textsc{LLM Dreaming} improves collaboration feasibility beyond capability matching.
RQ3 asks whether MapScore explanations faithfully identify decision-relevant matching factors.
RQ4 asks how \model{} compares with a negotiation-style agentic baseline.

\subsection{Setup}

\paragraph{Protocol.}
All experiments use the same suite of 20 matching tasks. Unless otherwise stated, capability outcomes use noise-free oracle observations. RQ1 uses up to 25 rounds, while RQ2 uses up to 30 rounds with consensus defined as three consecutive selections of the same candidate. The LLM component in \textsc{LLM Dreaming} is instantiated with \textbf{GPT-4o}.

\paragraph{Methods.}
We evaluate UCB + \(S_{\mathrm{cap}}\) Greedy, \(S_{\mathrm{cap}}\) Greedy only, and Random matching (mean over seeds).
Dreaming ablation compares \textsc{Greedy} and \textsc{Greedy + LLM Dreaming}.
For benchmarking we additionally compare with an AgenticPay-style negotiation baseline.
This baseline adapts the negotiation setting of AgenticPay~\citep{liu2026agenticpay} to collaboration matching, with additional motivation from multi-user coordination settings such as PeopleJoin~\citep{jhamtani-etal-2025-llm}.

\paragraph{Metrics.}
\label{met:metrics}
We report quality and efficiency metrics that separate match quality from online search cost.
For quality, we compute the oracle optimum $M_\text{optimum}$ from ground-truth capabilities using the fixed evaluation weight vector $\theta=[0.4,-0.1]$.
We report three recovery measures: $\rho_{\text{mode}}=M(\text{mode candidate},\text{true caps})/M_\text{optimum}$, $\rho_{\text{last}}=M(\text{last candidate},\text{true caps})/M_\text{optimum}$, and Hit rate, the fraction of tasks in which the analytical optimum is selected before the round budget expires.
For efficiency, we report the first-optimum round, the first consensus round at which the same candidate is selected $K=3$ consecutive times, and rounds-to-threshold metrics for $\rho\geq0.95$ and $\rho\geq0.99$.
For explanation faithfulness, we report perturbation-based score and decision effects, together with oracle-alignment metrics that compare the top-attributed factor with the strongest single-factor intervention.
Events that are not reached are capped at the corresponding round budget.

Inspired by AgenticPay \cite{liu2026agenticpay}, we investigated in the effect of users' personality in reaching consensus in LLM dreaming and compared it with pure Greedy selection on capability matches. We built a look-up table that contains the Global Score we obtained by evaluating AgenticPay on different combinations of personalities: Cooperative, Competitive, Analytical, Agreeable and Strategic. Here we adapted Global Score defined in \cite{liu2026agenticpay} to
$$
G = d \cdot(D + Q\cdot W + E)
$$
where $d=\gamma^{t-1}$, discount factor $\gamma$; negotiation success reward $D$, negotiation quality reward $W$, round efficiency reward $E$. 
$Q=4\cdot r_b\cdot r_s$ where $r_b \leftarrow (p^{max}-p) / Z$ and $r_s \leftarrow (p - p^{min}) / Z$, $p=$price, $p^{max}=$ candidate's max commitment score and $p^{min}=$ requester's min commitment score. Default values for the parameters can be found in Table \ref{tab:hyperparameters}. All variables inherit the same values from AgenticPay's default settings. We recorded the score in Table \ref{tab:personality_gt}. We compute an average top-3 compatibility (TKC) score under the look-up table:
\begin{equation}
\mathrm{TKC}(u) = \frac{1}{3} \sum_{v \in \text{top-3}(u)} C^*(p_u, p_v).
\end{equation}
where $C^*(p_u, p_v)$ is the look-up table score between a candidate with personality $p_u$ and a requester with personality $p_v$.
We additionally report Dream Score $S_{\text{sim}}$ - the LLM judge's
self-rated compatibility- to diagnose calibration. Note that the scores are calculated based on the separate AgentPay datasets\cite{liu2026agenticpay}, therefore the table is completely unknown to our model.

\paragraph{Benchmark.}
\label{sec:benchmark}
The evaluation set is fully synthetic and was generated by the simulator rather than collected from real users or manually annotated examples. The set was produced with a fixed random seed (42) and the mock backend that sampled from manually specified pools and ranges of the following traits: task titles, domains, role labels, skill vocabularies, capability values, candidates' and requesters' needs, requester profiles, candidate profiles, candidate cards, and latent context variables in the generation script, along with candidate capabilities centered around each task's required skills plus random perturbations. We then assigned each task's candidates evenly to four pre-specified experience tiers and add $\mathbf{Cap_{prior}}(\mu_{init}, \sigma_{init})$ while leaving the original candidate capabilities as ground truth. In total, the benchmark comprises 20 tasks, each pairing one requester with a pool of 20 candidates, yielding $20\times20=400$ requester–candidate pairs: the 20 tasks constitute independent matching episodes used as the paired unit for our significance tests, while the 400 pairs are the match evaluations underlying the aggregate quality and efficiency metrics.

\subsection{Online Selection}
To answer RQ1, we first isolate the analytical selector without \textsc{LLM Dreaming}.
This setting contains 20 tasks with 20 candidates per task, yielding 400 task--candidate matches (Section \ref{sec:benchmark}).
The oracle condition sets each prior mean to the true candidate mean, corresponding to zero prior noise.
Under a 25-round budget, the standard UCB warmup ("try each unseen candidate once" bonus) already consumes most of the interaction budget when there are 20 candidates,leaving limited opportunity for exploitation.

After adding tiered capabilities standard deviations and new candidate match score bonus, we observed that (1) Exploration until consensus decreased resulted in less exploration on newcomers but faster commit, (2) number of rounds to the first optimal pick decreased from 25 to around 11. (3) Before adding tiered standard deviation, quality of UCB + $S_\mathrm{cap}$ Greedy is $\rho_{\text{mode}}=0.869$. After adding tiered capabilities standard deviations, quality of UCB + $S_\mathrm{cap}$ Greedy is much closer to the analytical best $S_\mathrm{cap}$ Greedy -- $\rho_{\text{mode}}=0.983$ for $S_\mathrm{cap}$ Greedy vs $0.953$ for UCB + $S_\mathrm{cap}$ Greedy (Table \ref{tab:prior_sensitivity}). 

\begin{figure}[h!]
    \centering
    \includegraphics[width=0.4\textwidth]{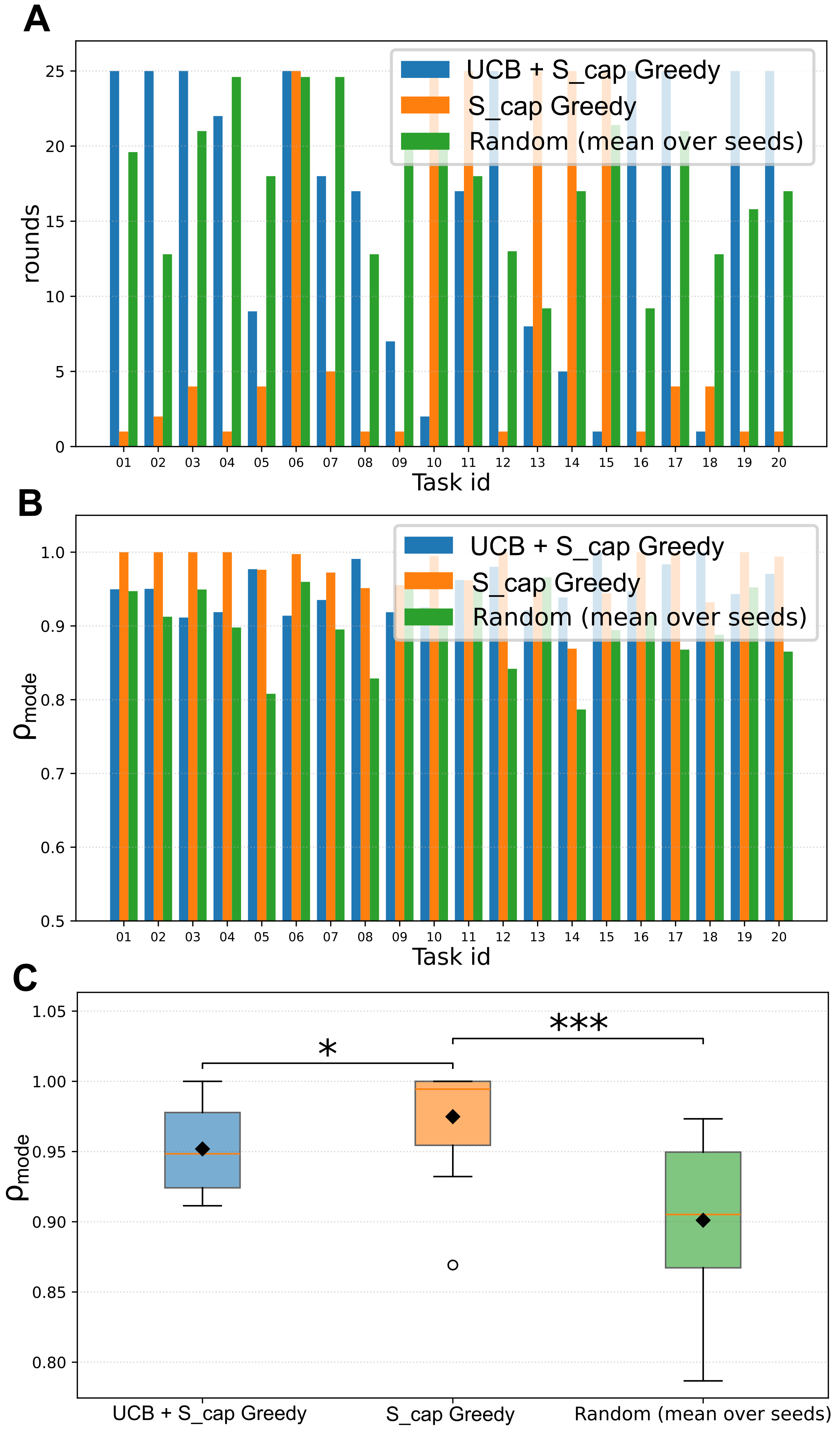}
    \caption{Selection on 20 matching tasks shows that tiered capability uncertainty improves both matching efficiency and quality under a limited round budget. A) Rounds to first optimal pick. B) Matching quality measured by $\rho_\text{mode}$. C) $\rho_\text{mode}$ distribution across tasks. P-values are calculated by Mann-Whitney U Test and are corrected by sample size. * means p-value $\leq$ 0.05 and *** means p-value $\leq$ 0.001.}
    \label{fig:20-online-learning}
\end{figure}

We also observed that our approach of combining UCB + $S_\mathrm{cap}$ Greedy exceeds $S_\mathrm{cap}$ Greedy-only mechanism on 6 of the 20 tasks and exceeds the random mechanism - mean selection over 5 seeds - on 10 out of 20 tasks (Figure \ref{fig:20-online-learning} A). In terms of quality, we reported $\rho_\text{mode}$ on all 20 tasks and observed that UCB + $S_\mathrm{cap}$ Greedy again exceeds $S_\mathrm{cap}$ Greedy-only on 6/20 tasks and exceeds random on 15/20 (Figure \ref{fig:20-online-learning} B). The distribution of $\rho_\text{mode}$ from UCB + $S_\mathrm{cap}$ Greedy is much closer to $S_\mathrm{cap}$ Greedy with P-value = 0.049.

\subsection{Effect of LLM Dreaming}
To answer RQ2, we evaluate whether \textsc{LLM Dreaming} improves process-level feasibility once the analytical selector is already strong.
The ablation uses the same 20 tasks and online-loop protocol, with at most 30 rounds per task and consensus declared after three consecutive selections of the same candidate.
The capability oracle is used only for evaluation.
We compare two variants: \textsc{Greedy}, which selects by analytical MapScore, and \textsc{Greedy + LLM Dreaming}, which reranks the shortlist using simulation compatibility and time-aware feasibility features.

\begin{table*}[t]
\centering
\scriptsize
\setlength{\tabcolsep}{4pt}
\begin{tabular}{@{}lcccccccc@{}}
\toprule
\textbf{Method}
  & $\rho_{\mathrm{mode}}$
  & $\rho_{\mathrm{last}}$
  & Hit$_{\mathrm{mode}}$
  & Hit$_{\mathrm{last}}$
  & $C_{\mathrm{mode}}$
  & $T_{\mathrm{mode}}$
  & Usable
  & $U_{\mathrm{comb}}$ \\
\midrule
\textsc{Greedy} & 0.9763 & 0.9730 & 0.35 & 0.30 & 0.6145 & 0.5862 & 0.45 & 0.8649 \\
\textsc{Greedy + LLM Dreaming} & 0.9575 & 0.9595 & 0.15 & 0.15 & 0.8000 & 0.7400 & 0.90 & 0.9042 \\
\bottomrule
\end{tabular}
\caption{Dreaming ablation under the official online-loop protocol.  $\rho_{\mathrm{mode}}$ and $\rho_{\mathrm{last}}$ are normalized true MatchScore values for the modal and final selected candidates.  Hit rates measure exact recovery of the oracle candidate.  $C_{\mathrm{mode}}$ is the Dreaming collaboration score, $T_{\mathrm{mode}}$ is the deterministic time-match score, Usable is the fraction of tasks whose selected candidate exceeds the time-feasibility threshold, and $U_{\mathrm{comb}}$ is combined utility.}
\label{tab:dreaming_ablation}
\end{table*}

Table~\ref{tab:dreaming_ablation} shows that analytical \textsc{Greedy} remains strongest on capability-oriented outcomes: it has higher $\rho_{\mathrm{mode}}$, $\rho_{\mathrm{last}}$, and higher oracle hit rates than \textsc{Greedy + LLM Dreaming}.  This is expected because MapScore directly optimizes the same capability signal as the oracle.  At the same time, \textsc{Greedy + LLM Dreaming} substantially improves collaboration, time feasibility, and combined utility, indicating that Dreaming contributes mainly to process-level feasibility while introducing some capability tradeoff.  We therefore treat Dreaming as a feasibility-aware correction layer rather than a replacement for analytical MapScore.

\subsection{Faithfulness Evaluation of MapScore Explanations}
\label{sec:faithfulness-eval}

To answer RQ3, we evaluate whether the factors surfaced by the MapScore decomposition are faithful to the analytical matching decision.
For each task, we rank candidates using deterministic MapScore, explain the top-ranked candidate, and perturb factors from its explanation. For a factor \(f\), we define the deletion effect as
\[
\Delta(f)=M(x)-M(x_{\setminus f}),
\]
where requirement factors zero the linked candidate capability and need factors zero the linked task offer. We report the resulting score drop and whether the original winner remains Top-1 after full-pool re-ranking.

We compare the top-attributed factor with a random positive factor and a single-factor perturbation oracle:
\[
f^\star_{\mathrm{oracle}}=\arg\max_f \Delta(f),
\;
\mathrm{ORat}
=
\frac{\Delta(f_{\mathrm{top}})}
{\Delta(f^\star_{\mathrm{oracle}})+\epsilon}.
\]
The oracle enumerates atomic interventions over winner capabilities and task offers. We also report OMatch, whether the explanation and oracle select the same intervention, and ORank, the rank of \(f_{\mathrm{top}}\) among all oracle interventions.

We evaluate four conditions: \textsc{Tiered Original}, the original 20-task benchmark; \textsc{Contrast Benchmark}, a less saturated diagnostic set with diversified candidate archetypes and randomized slots; \textsc{Contrast + SBERT}, the same contrast set with the SBERT/BGE encoder; and \textsc{Failure Case}, a controlled set with single off-topic needs and no directly matching offers.

\begin{table*}[t]
\centering
\scriptsize
\setlength{\tabcolsep}{3.6pt}
\renewcommand{\arraystretch}{0.92}
\begin{tabular}{lccccccccc}
\toprule
Condition
& Sat.$\downarrow$
& Gap$\uparrow$
& TDrop$\uparrow$
& RDrop
& TFlip$\uparrow$
& RFlip
& ORat$\uparrow$
& OMatch$\uparrow$
& ORank$\downarrow$
\\
\midrule
Tiered Original
& 0.94 & 0.021 & 0.528 & 0.150 & 1.00 & 0.95 & 0.981 & 0.95 & 1.05 \\
Contrast Benchmark
& 0.07 & 0.071 & 0.368 & 0.155 & 0.95 & 0.65 & 1.000 & 1.00 & 1.00 \\
Contrast + SBERT
& 0.08 & 0.065 & 0.347 & 0.152 & 0.95 & 0.70 & 1.000 & 0.95 & 1.05 \\
Failure Case
& 0.04 & 0.140 & 0.092 & 0.171 & 0.10 & 0.50 & 0.291 & 0.10 & 3.70 \\
\bottomrule
\end{tabular}
\caption{
Faithfulness evaluation. Sat. is the fraction of candidates with saturated \(S_{\mathrm{cap}}\); Gap is the winner-runner-up score gap. TDrop/RDrop and TFlip/RFlip compare top-attributed and random-factor perturbations. ORat, OMatch, and ORank compare the top-attributed factor with the strongest single-factor oracle intervention.
}
\label{tab:faithfulness}
\end{table*}

Table~\ref{tab:faithfulness} shows that attribution is most informative on the contrast benchmark: top-attributed factors outperform random factors in both score drop and ranking flip, and match the strongest single-factor oracle intervention. The trend remains under SBERT/BGE, with only one oracle mismatch across 20 tasks. The tiered benchmark mainly serves as a saturation diagnostic.

The Failure Case gives the intended reversal: top-attributed factors no longer outperform random factors, and oracle agreement drops sharply. This is the expected boundary case. A single off-topic need can receive high score share through attention matching, while perturbing its linked offer may have little effect when attention moves to substitute offers. We provide the derivation and examples in Appendix~\ref{app:faithfulness-details}.

Finally, we repeat random and shuffled baselines over five seeds on the Contrast Benchmark. The task set is fixed, and the seed controls factor sampling. The main metrics remain stable: RDrop is \(0.1673\pm0.0131\), RFlip is \(0.6800\pm0.0748\), and ORat remains \(1.0000\).

\subsection{Benchmark \model against human-agent networks}
To answer RQ4, we compare \model{} with an AgenticPay \cite{liu2026agenticpay} negotiation baseline, as well as PeopleJoin \cite{peoplejoin} on 20 tasks with 3 candidates per task. For AgenticPay, we switched the role of seller to requester and the role of buyer to candidates, and trading deals to collaboration negotiation tasks. Then we asked it to match requester and candidates one by one to match the requester with the best candidate by computing quality score of the negotiation \cite{liu2026agenticpay}. We evaluated both methods on both matching quality and efficiency metrics (Methods \ref{met:metrics}). We observed that \model exceeds AgenticPay on all metrics (Table \ref{tab:benchmark}).

\subsection{Personality Compatibility Evaluation}
The MapScore selector in \model is capability-driven and explicitly \emph{personality-blind}. We hypothesize that adding LLM dreaming as a re-ranking stage allows the system to identify personality-driven collaboration risks that capability scoring cannot capture. We use a pre-computed $5{\times}5$ Global Score matrix from \textsc{AgenticPay} \cite{liu2026agenticpay}, an independent structured negotiation simulator, as held-out ground truth for personality compatibility (Table \ref{tab:personality_gt}). We construct 25 matching tasks: 5 requester personalities $\times$ 5 task instances per personality, yielding 5 tasks per personality and 25 tasks in total. Each task consists of one requester and a candidate pool of size~5, with one candidate per personality. To isolate personality-driven selection, all candidates have identical capability vectors so that the analytical MapScore is constant across the pool. We compare two selection mechanism: \textsc{Greedy}: ranks candidates by analytical MapScore $M$; \textsc{Dreaming}: simulates a three-turn agent-agent negotiation for each candidate using \textbf{GPT-4.1}, then ranks by a weighted combination of $M$ and the LLM-judged compatibility, e.g.$S_{\text{sim}}$ (50/50). The top-3 ranked candidates form the final selection (Table \ref{tab:per-personality-comparison}). Dreaming improves over Greedy on 3/5 = 60\% personalities in terms of TKC, and 4/5 = 80\% personalities in terms of $S_\mathrm{sim}$. Dreaming surfaces personality signal that capability scoring cannot. On 3 of 5 requester personalities, Dreaming improves top-3 compatibility by $+0.042$ to $+0.071$ (Table \ref{tab:per-personality-comparison}). The strongest case is Competitive ($\mathrm{TKC} = 0.988$). The disagreement between Dreaming and Greedy is the strongest when requester personality is Agreeable. This is because LLM dreaming only detects agreement but does not detect a \textit{fair} agreement - an agreeable requester agrees with candidates immediately but the requester is actually exploited (Figure \ref{fig:agreeable_trap}).

\section{Conclusion}
\label{sec:conclusion}
We studied collaboration matching in mixed human-agent networks as a task-conditioned problem of bidirectional fit rather than a static ranking of globally strong collaborators. We introduced \model{}, which represents tasks and candidates through explicit needs, offers, and capabilities, scores matches with feasibility-gated MapScore, updates selections through online exploration, uses \textsc{LLM Dreaming} to account for process-level feasibility, and exposes decision factors through a closed-form explanation decomposition. Across the evaluation, analytical online selection recovered high-quality collaborators under limited interaction budgets, while \textsc{LLM Dreaming} improved collaboration and time feasibility even when it introduced a small capability-optimality tradeoff. Perturbation-based diagnostics further showed that the highlighted MapScore factors align with decision-relevant interventions in the main contrast setting. Compared with an AgenticPay-style negotiation baseline, \model{} achieved stronger quality and efficiency in the adapted collaboration benchmark. Our results establish a foundation for explicitly modeling bidirectional, feasibility-gated and uncertainty-aware collaboration. This mechanism is worth carrying into real-world scientific projects with human participants and observing long-term collaboration outcomes.

\section*{Limitations}
\label{sec:limitation}
Several limitations remain. First, our experiments use a controlled 20-task suite with finite candidate pools and, unless otherwise stated, noise-free oracle capability observations; this design isolates the matching mechanism but does not capture all uncertainty in real deployments, including noisy feedback, strategic behavior, and long-term relationship dynamics. Second, \model{} depends on explicit task and candidate representations, which make MapScore interpretable but may omit softer collaboration factors such as trust, communication style, availability, institutional constraints, and evolving preferences. Third, \textsc{LLM Dreaming} is currently used as a feasibility-aware reranking layer rather than a full structured social simulation. Prior work suggests that LLM agents with debate and reflection can exhibit richer collaborative behaviors such as consensus reaching and conformity~\citep{zhang2024exploring}, suggesting that future versions could stress-test requester--candidate fit with explicit roles such as a requester, candidate, constraint checker, and mediator. Finally, our AgenticPay-style comparison adapts a negotiation framework to collaboration matching, so the benchmark result should be interpreted as evidence against this adapted negotiation baseline rather than as a claim about all agentic matching systems. Future work should therefore evaluate \model{} in noisier deployments, with richer social simulation, broader baselines, and longer-horizon collaboration outcomes.

\section*{Author Contributions}
Jiayao Gu (Conceptualization[equal], Methodology[equal], Software[equal], Formal analysis[equal], Data Curation[Lead], Writing-Original Draft[equal], Writing-Review \& Editing[equal], Visualization[Lead],  Supervision[lead], Project administration [Lead]), Kexin Chu (Conceptualization[equal], Methodology[equal], Software[equal], Formal analysis[equal], Writing-Original Draft[equal], Writing-Review \& Editing[equal]), Peidong Liu (Methodology[equal], Software[equal], Formal analysis[equal], Writing - Original Draft[equal], Writing-Review \& Editing[equal]), Yue Yang (Funding acquisition[minor]), Lynn Ai (Funding acquisition[Lead]), Qi Zhang (Writing-Original Draft[minor]), Ling Yang (Conceptualization[minor]), Tianyu Shi (Conceptualization[Lead], Project administration[minor]).

\clearpage
\bibliography{ref}

\clearpage
\appendix
\renewcommand{\thetable}{S\arabic{table}}
\renewcommand{\thefigure}{S\arabic{figure}}
\counterwithin{figure}{section}
\counterwithin{table}{section}
\setcounter{figure}{0}
\setcounter{table}{0}

\section{Additional Background and Related Work}
\label{app:bg-related}

\subsection{Collaboration Matching as Bidirectional Fit}

Collaboration matching differs from standard recommendation because the value of a match is task-conditioned and bidirectional. A requester may need collaborators who cover concrete missing capabilities, but a candidate also needs the task to provide sufficient value, appropriate incentives, and feasible working conditions. A high-capability candidate is therefore not necessarily a good collaborator for a given task if the task does not satisfy the candidate's needs or violates hard constraints such as access, consent, availability, or conflict-of-interest requirements. This motivates modeling collaboration as a constrained fit between requester gaps, candidate offers, candidate needs, and task-level feasibility.

This view is especially important in mixed human-agent networks. Human researchers, proxy agents, service agents, and reusable skills may all be selectable candidates, but they differ in capability uncertainty, interaction cost, reliability, and permissible use. A useful matching system must therefore handle cold-start candidates, update uncertain estimates from feedback, and provide interpretable reasons for each recommendation. In \model{}, these requirements motivate bidirectional MapScore, hard feasibility gates, uncertainty-aware online exploration, and explanation decomposition.

\subsection{Relation to Existing Matching and Agent Systems}

Recommender systems provide a broad foundation for ranking items or people under user preferences, feedback, cold start, and platform objectives~\citep{adomavicius2005toward,bobadilla2013recommender}. Multi-stakeholder recommendation further emphasizes that recommendations can affect multiple parties with different utilities and fairness concerns~\citep{abdollahpouri2020multistakeholder,burke2025decentering}. These perspectives are relevant to collaboration formation, but they do not directly model whether a candidate fills a task-specific capability gap while also receiving sufficient participation value from the task.

Expert team formation and matching markets are also closely related. Team formation work studies how to assemble groups with appropriate skills and collaboration structure~\citep{lappas2009team,saeedi2025survey}, while stable and learned matching methods reason about two-sided preferences, stability, or uncertainty~\citep{gale1962college,li2025bandit}. \model{} differs by focusing on task-conditioned collaboration edges rather than global compatibility: each match is evaluated by residual requester needs, candidate-side value, hard feasibility constraints, and online uncertainty.

Recent LLM-agent work studies multi-agent coordination, human-agent collaboration, planning, and information gathering~\citep{huang-etal-2025-enable,jhamtani-etal-2025-llm,wang-etal-2025-anymac,sun-etal-2025-collab}. These systems show that LLM agents can support complex collaborative workflows, but they typically do not provide an explicit, learnable, and decomposable matching score for deciding which human, agent, or skill should be connected to a task. \model{} is complementary: it treats matching as the upstream decision problem that selects promising collaborators before downstream coordination occurs.

Work on tool use and efficient agent execution highlights another relevant dimension: agents often need to choose among functionally related capabilities under latency, reliability, and resource constraints~\citep{xu-etal-2025-learning,chu2026latency,chu2025mcam,chu2025selective}. In our setting, such constraints appear as part of candidate feasibility and process-level collaboration risk. The goal of \model{} is not to replace planning or tool-use policies, but to provide a task-conditioned matching layer that can explain, update, and filter collaboration candidates before expensive interaction or simulation.

\paragraph{Baseline landscape.}

Table~\ref{tab:baseline-landscape} summarizes the baseline landscape behind our evaluation design.
We did not find a prior system that directly implements task-conditioned scientific collaboration matching over mixed human-agent candidates.
Recommendation and multistakeholder RecSys provide ranking, exposure, and cold-start foundations, but do not directly model task-specific missing capabilities or feasibility-gated collaboration edges.
Expert team formation is the closest classical formulation for skill coverage and team structure, yet it assumes human expert pools and does not model candidate-side value, online uncertainty, or mixed human-agent candidates.
Two-sided and bandit matching markets provide preference-learning theory, but their preferences are not decomposed into semantic capability gaps, task offers, hard gates, and explanation factors.
LM-agent coordination systems study collaborative execution, teammate discovery, and communication protocols, but typically start after agents or users have been selected.
Among these, PeopleJoin~\citep{peoplejoin} is directly runnable as an LLM-agent baseline for our setting: it coordinates multi-user information gathering by identifying which users are likely to hold the requested information, which we adapt to collaboration matching by treating candidate profiles as the queryable knowledge sources and asking the coordinator to identify which candidate best fills a requester's task.
AgenticPay~\citep{liu2026agenticpay} provides the closest executable agentic \emph{negotiation} comparison: it supports multi-round negotiation with private constraints and valuations, and we adapt its buyer-seller interaction into requester-candidate collaboration for RQ4.
We therefore evaluate internal selector ablations for the components that can be isolated and use PeopleJoin and the AgenticPay-style negotiation as the two closest runnable comparisons, while treating the remaining rows as adjacent problem families rather than direct drop-in baselines.
In the compact matrix, Gap denotes capability-gap modeling, Value denotes two-sided value, Hard denotes hard feasibility constraints, and Online denotes online or cold-start support; \cmark{} marks a central objective, \pmark{} marks partial support, and \xmark{} marks a non-central criterion.

\begin{table}[H]
\scriptsize
\setlength{\tabcolsep}{7pt}
\renewcommand{\arraystretch}{1.12}
\begin{tabular}{p{0.34\linewidth}
>{\centering\arraybackslash}p{0.08\linewidth}
>{\centering\arraybackslash}p{0.08\linewidth}
>{\centering\arraybackslash}p{0.08\linewidth}
>{\centering\arraybackslash}p{0.08\linewidth}}
\toprule
\textbf{Approach} &
\textbf{Gap} &
\textbf{Value} &
\textbf{Hard} &
\textbf{Online} \\
\midrule
RecSys / multi-stake. & \pmark & \pmark & \pmark & \cmark \\
Team formation & \cmark & \xmark & \pmark & \xmark \\
Two-sided / bandit & \xmark & \cmark & \pmark & \cmark \\
LLM-agent coordination & \pmark & \xmark & \pmark & \xmark \\
PeopleJoin & \pmark & \xmark & \pmark & \xmark \\
AgenticPay-style & \xmark & \cmark & \cmark & \xmark \\
\textbf{\model{}} & \cmark & \cmark & \cmark & \cmark \\
\bottomrule
\end{tabular}
\caption{Baseline matrix. PeopleJoin and AgenticPay-style are the two runnable comparison baselines.}
\label{tab:baseline-landscape}
\end{table}

\clearpage
\section{Additional Method Details}
\label{app:method-details}

\subsection{Attention-Based Semantic Matching}
\label{app:semantic-matching}

We instantiate capability and offer matching with attention over open-set semantic items.
Task requirements are written as
\[
\mathbf Q_T=\{(\mathbf q_j,q_j,w_j^T)\}_{j=1}^{m},
\]
where $\mathbf q_j$ is a requirement embedding, $q_j$ is the required level, and $w_j^T$ is its task-specific importance.
Candidate needs and task offers are written as
\[
\mathbf{Need}_v=\{(\mathbf z_l^v,n_l^v)\}_{l=1}^{h_v},
\qquad
\mathbf O_T=\{(\mathbf o_k,o_k)\}_{k=1}^{r}.
\]

For each task requirement $\mathbf q_j$, requester-side coverage is computed by attention over requester capabilities:
\[
\begin{aligned}
\alpha_{i,j}^{u}
&=
\frac{\exp(\mathrm{sim}(\mathbf q_j,\mathbf e_i^u)/\tau)}
{\sum_k \exp(\mathrm{sim}(\mathbf q_j,\mathbf e_k^u)/\tau)},\\
\tilde p_u^j
&=\sum_i \alpha_{i,j}^{u}p_i^u .
\end{aligned}
\]
Candidate-side proficiency and uncertainty are computed analogously as $\tilde p_v^j$ and $\tilde\sigma_v^j$.
We use an optimistic coverage estimate
\[
c_v^j=\min(1,\tilde p_v^j+\beta_t\tilde\sigma_v^j),
\]
where $\beta_t$ controls uncertainty-aware optimism.
The capability-gap score is
\[
S_{\mathrm{cap}}(u,v,T)=
\frac{\sum_j w_j^T\min(\mathrm{Gap}_j(u,T),c_v^j)}
{\sum_j w_j^T\mathrm{Gap}_j(u,T)+\epsilon}.
\]

For participation-value fit, each candidate need attends over task offers:
\[
\begin{aligned}
\gamma_{k,l}^{v}
&=
\frac{\exp(\mathrm{sim}(\mathbf z_l^v,\mathbf o_k)/\tau_o)}
{\sum_{k'}\exp(\mathrm{sim}(\mathbf z_l^v,\mathbf o_{k'})/\tau_o)},\\
\hat o_l^T
&=\sum_k \gamma_{k,l}^{v}o_k .
\end{aligned}
\]

 To ensure mutual benefit, we quantify how well a task satisfies a researcher's needs. We first compute the benefit vector offered by task $T$ as $o_T^k = \alpha \cdot Q_T^k$, then define the need fulfillment score as 
 \[
 S_{\text{need}}(v, T) = \frac{\sum_{k=1}^{K} \min(o_T^k,, n_v^k) \cdot w_k}{\sum_{k=1}^{K} n_v^k \cdot w_k + \epsilon}, 
 \]
 which measures the weighted fraction of researcher $v$'s needs that the task can satisfy, capped at actual need levels to avoid over-counting. This score lies in $[0, 1]$ and guarantees that high-scoring matches are attractive to both parties.

\subsection{Closed-form of MapScore}
\label{app:mapscore_closed_form}
MapScore can be decomposed over the same explicit factors used for matching. For each feasible match, we attribute the final score to requester-side requirement contributions and candidate-side need contributions.

For the capability term, let \(a_j\) be the normalized importance of requirement \(j\), and let \(\mathrm{Gap}_j(u,T)\) be the residual capability gap of requester \(u\) on task \(T\). Given candidate \(v\)'s soft coverage \( c_v^j\) of requirement \(j\), the useful coverage is capped by the requester gap:
\[
\mathrm{Cover}_j(u,v,T)
=
\min\left(c_v^j,\mathrm{Gap}_j(u,T)\right).
\]
The capability score is
\[
S_{\mathrm{cap}}(u,v,T)
=
\frac{
\sum_j a_j \mathrm{Cover}_j(u,v,T)
}{
\sum_j a_j \mathrm{Gap}_j(u,T)+\epsilon
}.
\]
Therefore, the contribution of requirement \(j\) to the final MapScore is
\[
C^{\mathrm{req}}_j(u,v,T)
=
\sigma(u,v,T)\lambda
\frac{
a_j \mathrm{Cover}_j(u,v,T)
}{
\sum_k a_k \mathrm{Gap}_k(u,T)+\epsilon
}.
\]
This term measures how much the match score depends on candidate \(v\)'s coverage of a specific requester-side gap.

For the need term, let \(\rho_\ell\) be the intensity of candidate need \(\ell\), and let \(\tilde o_\ell\) be the attention-weighted value provided by task offers. The satisfied need value is
\[
\mathrm{Sat}_\ell(v,T)
=
\min\left(\tilde o_\ell,\rho_\ell\right).
\]
The need score is
\[
S_{\mathrm{need}}(v,T)
=
\frac{
\sum_\ell \mathrm{Sat}_\ell(v,T)\rho_\ell
}{
\sum_\ell \rho_\ell^2+\epsilon
}.
\]
The corresponding need-level contribution is
\[
C^{\mathrm{need}}_\ell(u,v,T)
=
\sigma(u,v,T)(1-\lambda)
\frac{
\mathrm{Sat}_\ell(v,T)\rho_\ell
}{
\sum_k \rho_k^2+\epsilon
}.
\]
This term measures how much the score depends on task \(T\)'s satisfaction of a specific candidate-side participation need.

For each feasible pair \((u,v,T)\), we collect all positive contribution terms
\[
\mathcal{F}^{+}(u,v,T)
=
\{C^{\mathrm{req}}_j>0\}_j
\cup
\{C^{\mathrm{need}}_\ell>0\}_\ell ,
\]
and define the top-attributed factor as
\[
f_{\mathrm{top}}
=
\arg\max_{f\in\mathcal{F}^{+}} C_f.
\]
The explanation returned with a match consists of the dominant covered requirements, the dominant satisfied needs, the top-attributed factor, and the feasibility gate status.

\subsection{Feedback Update Equations}
\label{app:feedback-updates}

When feedback provides evidence about a capability item, we update its proficiency and uncertainty with a Gaussian posterior update.
For an observation $x_t$ with noise variance $\sigma_{\mathrm{obs}}^2$ and prior $(\mu_{i,t}^v,(\sigma_{i,t}^v)^2)$,
\[
\mu_{i,t+1}^v=
\frac{\mu_{i,t}^v/(\sigma_{i,t}^v)^2+x_t/\sigma_{\mathrm{obs}}^2}
{1/(\sigma_{i,t}^v)^2+1/\sigma_{\mathrm{obs}}^2},
\]
\[
(\sigma_{i,t+1}^v)^2=
\left(
\frac{1}{(\sigma_{i,t}^v)^2}
+\frac{1}{\sigma_{\mathrm{obs}}^2}
\right)^{-1}.
\]

We update the balance parameter $\lambda$ by minimizing prediction loss between observed collaboration outcome $M$ and predicted map score $\hat M$:
\[
\begin{aligned}
\mathcal L_t
&=(M - \hat{M})^2,\\
\hat M
&=\lambda S_{\mathrm{cap}}(u,v,T) +(1-\lambda)S_{\mathrm{need}}(v,T)\\
\theta &\leftarrow \theta - \eta\nabla_\theta L - \lambda(\theta - \theta_{prior})
\end{aligned}
\]

\subsection{Explanation Faithfulness Protocol and Diagnostics}
\label{app:faithfulness-details}

\paragraph{Scoring and perturbation protocol.}
Faithfulness tests use deterministic MapScore with exploration disabled. For each task, we rank all candidates, explain the top-ranked candidate \(v^\star\), and evaluate interventions on its positive explanation factors \(\mathcal{F}^{+}(u,v^\star,T)\) defined in Section~3.4. We use the largest-contribution factor as \(f_{\mathrm{top}}\), sample \(f_{\mathrm{rand}}\) from the same positive factor set, and use the smallest positive factor only for supplementary diagnostics.

For a factor \(f\), the deletion effect is
\[
\Delta(f)=M(x)-M(x_{\setminus f}),
\]
where \(x_{\setminus f}\) weakens the primitive linked to \(f\). A requirement factor zeros the linked candidate capability proficiency, \(p_i^{v^\star}\leftarrow 0\); a need factor zeros the linked task offer strength, \(s_r\leftarrow 0\). After each perturbation, we recompute MapScore for the full candidate pool and record both the score drop and whether the original winner remains Top-1.

\paragraph{Single-factor perturbation oracle.}
Random factors test whether the explanation is better than arbitrary factor selection. We additionally use a stricter single-factor oracle. Let
\[
\begin{aligned}
\mathcal{I}(u,v^\star,T)
&=
\{p_i^{v^\star}\leftarrow 0:
i\in\mathrm{Cap}_{v^\star}\}\\
&\quad\cup
\{s_r\leftarrow 0:r\in O_T\}.
\end{aligned}
\]
be the set of atomic interventions available to the explanation. The oracle intervention is
\[
f^\star_{\mathrm{oracle}}
=
\arg\max_{f\in\mathcal{I}(u,v^\star,T)} \Delta(f).
\]
We report three oracle metrics. First,
\[
\mathrm{ORat}
=
\frac{
\Delta(f_{\mathrm{top}})
}{
\Delta(f^\star_{\mathrm{oracle}})+\epsilon
}.
\]
Second, OMatch checks whether the explanation and the oracle select the same atomic intervention:
\[
\mathrm{OMatch}
=
\mathbb{I}
\left[
\pi(f_{\mathrm{top}})
=
\pi(f^\star_{\mathrm{oracle}})
\right],
\]
where \(\pi(f)\) maps an explanation factor to its linked capability or offer intervention. Third, ORank is the rank of the explanation factor among all oracle-enumerated interventions:
\[
\begin{aligned}
\mathrm{ORank}
&=
1+\\
&\quad\bigl|
\{f\in\mathcal{I}(u,v^\star,T): \Delta(f)>\Delta(f_{\mathrm{top}})\}
\bigr|.
\end{aligned}
\]

\paragraph{Diagnostic benchmark construction.}
The original tiered benchmark is useful for evaluating matching quality, but it is often saturated for capability explanations: many candidates already cover most requester-side gaps. We therefore construct a contrastive faithfulness benchmark at the same 20-task scale. Each task contains candidates drawn from several archetypes, summarized in Table~\ref{tab:contrast-archetypes}. Candidate slots are randomized, and the intended winner archetype is recorded in task metadata. This prevents the diagnostic from depending on a fixed candidate position or a single winner profile.

\begin{table*}[t]
\centering
\small
\setlength{\tabcolsep}{4pt}
\begin{tabular}{lll}
\toprule
Archetype & Capability profile & Need profile \\
\midrule
High-capability / low-need
& Strong gap coverage
& Weak task-side value \\
Low-capability / high-need
& Weak gap coverage
& Strong task-side value \\
Balanced
& Moderate coverage
& Moderate value \\
Specialist
& Strong on one key gap
& Task-dependent value \\
Distractor
& Off-target coverage
& Task-dependent value \\
\bottomrule
\end{tabular}
\caption{
Candidate archetypes used in the contrastive faithfulness benchmark. The benchmark is designed to reduce \(S_{\mathrm{cap}}\) saturation and diversify winner profiles.
}
\label{tab:contrast-archetypes}
\end{table*}

The failure benchmark contains controlled tasks designed to expose a boundary of share-based attribution. Each candidate has a single high-intensity need that is off-topic with respect to the task offers. The task offers contain no direct semantic counterpart to that need. This setting encourages replaceable attention links: the explanation may assign high contribution to the need through its current matched offer, while perturbing that offer may simply move attention to other weakly related offers.

\paragraph{Why the failure case breaks share-based attribution.}
The failure mechanism follows from the capped need satisfaction term. For a candidate with a single need of intensity \(\rho\),
\[
\begin{aligned}
S_{\mathrm{need}}(v,T)
&=
\frac{
\rho\,\mathrm{Sat}
}{
\rho^2+\epsilon
}
\approx
\frac{\mathrm{Sat}}{\rho},\\  
\mathrm{Sat}&=\min(\tilde{o},\rho).
\end{aligned}
\]
When the attention-weighted offer value reaches the cap, \(\tilde{o}\ge\rho\), we have \(\mathrm{Sat}=\rho\), so \(S_{\mathrm{need}}\approx 1\). Since this is the only need, its contribution can dominate the need-side score allocation.

Now suppose the need is linked to an offer \(r^\star\) only through a weak semantic match. The matched offer value is
\[
\tilde{o}
=
\sum_r \beta_r s_r .
\]
After perturbing the linked offer,
\[
s_{r^\star}\leftarrow 0,
\qquad
\tilde{o}'
=
\sum_{r\ne r^\star}\beta'_r s_r ,
\]
where the attention weights are recomputed. If substitute offers still satisfy the cap,
\[
\tilde{o}'\ge\rho,
\]
then
\[
\begin{aligned}
\mathrm{Sat}'&=\rho,\\ 
S_{\mathrm{need}}'(v,T)
&\approx S_{\mathrm{need}}(v,T),\qquad
\Delta(f)\approx 0.
\end{aligned}
\]
Thus a factor can receive high contribution share under the current decomposition while having weak single-factor perturbation effect. This is the boundary tested by the Failure Case in Table~\ref{tab:faithfulness}.

\paragraph{Stochastic baseline stability.}
The multi-seed analysis in Section~4.4 repeats only the random and shuffled factor sampling. The task set and candidate profiles are fixed. Across five seeds on the Contrast Benchmark, the main stochastic metrics remain stable:
\[
\begin{aligned}
\mathrm{RDrop}&=0.1673\pm0.0131,\\
\mathrm{RFlip}&=0.6800\pm0.0748,\\
\mathrm{ORat}&=1.0000.
\end{aligned}
\]
This confirms that the oracle-level agreement does not depend on a favorable random-factor draw.

\subsection{Computational Resources}
Evaluation runs issued approximately 480 API calls to \texttt{openai/gpt-4.1} and \texttt{openai/gpt-4o}, consuming an estimated 480\,000 tokens and incurring a cost of roughly \$1.20--\$1.60 at list pricing, with a wall-clock runtime of 40--60 minutes per run executed sequentially.

\begin{table*}[h]
\centering
\label{tab:tasks}
\small
\begin{tabular}{@{}cllp{7.2cm}@{}}
\toprule
\textbf{ID} & \textbf{Title} & \textbf{Domain} & \textbf{Required Skills} \\
\midrule
\texttt{T01} & Design Recommendation API & Ml Platform & ml systems, python, evaluation, paper writing, bayesian modeling \\
\texttt{T02} & Launch RAG Evaluation Dashboard & Industry & bayesian modeling, recommender systems, communication \\
\texttt{T03} & Create ML Data Pipeline & Industry & frontend, bayesian modeling, python \\
\texttt{T04} & Prototype Hiring Assistant & Analytics & backend, recommender systems, communication, statistics \\
\texttt{T05} & Improve Matching Quality Monitor & Research & bayesian modeling, backend, evaluation, frontend \\
\texttt{T06} & Ship Research Collaboration Portal & Industry & recommender systems, bayesian modeling, product design, paper writing \\
\texttt{T07} & Develop Demand Forecasting Workflow & Matching & data analysis, communication, frontend, statistics, paper writing \\
\texttt{T08} & Build Knowledge Graph Explorer & Analytics & statistics, bayesian modeling, ml systems \\
\texttt{T09} & Optimize Retrieval Stack & Industry & recommender systems, data analysis, python \\
\texttt{T10} & Create Project Risk Scorer & Ml Platform & communication, evaluation, paper writing, recommender systems \\
\texttt{T11} & Develop User Segmentation Toolkit & Ml Platform & evaluation, data analysis, ml systems, recommender systems \\
\texttt{T12} & Deploy Online Learning Service & Industry & backend, bayesian modeling, paper writing \\
\texttt{T13} & Design Task Allocation Engine & Matching & communication, product design, recommender systems, bayesian modeling, distributed systems \\
\texttt{T14} & Implement Experiment Tracking UI & Ml Platform & python, data engineering, frontend \\
\texttt{T15} & Build Collaboration Outcome Predictor & Research & frontend, backend, bayesian modeling \\
\texttt{T16} & Create Portfolio Ranking Service & Industry & backend, paper writing, data engineering, python, frontend \\
\texttt{T17} & Prototype Industry Search & Matching & backend, data analysis, statistics \\
\texttt{T18} & Develop Causal Analysis Notebook & Analytics & paper writing, data analysis, communication, evaluation, python \\
\texttt{T19} & Ship Candidate Vetting Tool & Research & data analysis, communication, evaluation, distributed systems \\
\texttt{T20} & Build Bayesian Churn Model & Matching & evaluation, communication, paper writing, statistics, project management \\
\bottomrule
\end{tabular}
\caption{Overview of the 20 benchmark tasks. Each task is drawn from a distinct initiative domain; required skills are the competencies that candidates must cover.}
\end{table*}

\begin{table}[b]
\centering
\small
\setlength{\tabcolsep}{5pt}
\begin{tabular}{@{}l cc@{}}
\toprule
\textbf{Method}
  & First Opt.\ $\downarrow$
  & $\rho_\mathrm{mode}$ $\uparrow$ \\
\midrule
\multicolumn{3}{@{}l}{\textit{Uniform priors} ($\sigma_\mathrm{init}=0.25$)} \\[2pt]
UCB + $S_\text{cap}$ Greedy        & $11.0 \pm 5.9$  & $0.896 \pm 0.063$ \\
$S_\mathrm{cap}$ Greedy & $\mathbf{1.0 \pm 0.0}$ & $\mathbf{0.982 \pm 0.021}$ \\
Random            & $17.6 \pm 4.7$  & $0.901 \pm 0.053$ \\
\midrule
\multicolumn{3}{@{}l}{\textit{Tiered priors} (tier-stratified $\mu_\mathrm{init}, \sigma_\mathrm{init}$)} \\[2pt]
UCB + $S_\text{cap}$ Greedy        & $16.6 \pm 9.3$  & $0.952 \pm 0.029$ \\
$S_\mathrm{cap}$ Greedy & $\mathbf{9.1 \pm 10.5}$  & $\mathbf{0.975 \pm 0.033}$ \\
Random            & $17.6 \pm 4.7$  & $0.901 \pm 0.053$ \\
\bottomrule
\end{tabular}
\caption{%
  Effect of prior type on matching quality and efficiency over 20 tasks
  (20 candidates each, $N_\mathrm{max}=25$ rounds, no noise).
  \emph{Uniform priors}: all candidates initialised with $\sigma_\mathrm{init}=0.25$.
  \emph{Tiered priors}: candidates initialised with tier-stratified $(\mu_\mathrm{init}, \sigma_\mathrm{init})$
  derived from ground-truth capability tiers.
  Values are mean $\pm$ std across 20 tasks.
  $\downarrow$ lower is better; $\uparrow$ higher is better.%
}
\label{tab:prior_sensitivity}
\end{table}

\begin{table*}[t] 
\centering 
\small 
\setlength{\tabcolsep}{6pt}  
\begin{tabular}{@{}llccc@{}} 
\toprule 
\textbf{Dimension} & \textbf{Metric} & \textbf{AgenticPay} & \textbf{PeopleJoin} & \textbf{\model (ours)} \\ 
\midrule \multirow{3}{*}{\textit{Quality} $\uparrow$} & $\rho_{\mathrm{mode}}$ & $0.941 \pm 0.066$ & $0.966 \pm 0.042$ & $\mathbf{0.977 \pm 0.046}$ \\ 
& $\rho_{\mathrm{last}}$ & $0.941 \pm 0.066$ & $0.966 \pm 0.042$ & $\mathbf{0.989 \pm 0.022}$ \\ 
& Hit Rate & $0.250 \pm 0.444$ & $0.333 \pm 0.471$ & $\mathbf{0.750 \pm 0.444}$ \\ 
\midrule \multirow{4}{*}{\textit{Efficiency} $\downarrow$} & First Opt.\ (rounds) & $20.0 \pm 0.0$ & $8.38 \pm 2.32$ & $\mathbf{4.4 \pm 7.2}$ \\ 
& Consensus (rounds) & $20.0 \pm 0.0$ & $\mathbf{5.10 \pm 0.60}$ & $6.5 \pm 4.3$ \\ 
& Rounds to $\rho \geq 0.95$ & $20.0 \pm 0.0$ & $5.14 \pm 0.60$ & $\mathbf{1.61 \pm 1.20}$ \\ 
& Rounds to $\rho \geq 0.99$ & $20.0 \pm 0.0$ & $5.21 \pm 0.66$ & $\mathbf{2.06 \pm 1.73}$ \\ 
\bottomrule 
\end{tabular} 
\caption{Benchmark results comparing CoWeaver, PeopleJoin-Reactive, and AgenticPay. Quality metrics (higher is better) and efficiency metrics (lower is better) are reported as mean $\pm$ std. PeopleJoin, AgenticPay and CoWeaver are evaluated using the 20-task protocol. For rounds-to-threshold metrics, parentheses indicate the number of task trajectories that reached the corresponding threshold. The best result for each metric is shown in bold. CoWeaver achieves the highest matching quality and reaches high-quality candidates in the fewest rounds, while PeopleJoin exhibits faster repeated-selection consensus. }
\label{tab:benchmark} 
\end{table*}

\begin{table*}[h]
\centering
\small
\begin{tabular}{l|ccccc}
\toprule
Candidate $\setminus$ Requester & Cooperative & Competitive & Analytical & Agreeable & Strategic \\
\midrule
Cooperative & 0.57 & 0.46 & 0.87 & 0.98 & 0.56 \\
Competitive & 0.98 & 0.00 & 0.99 & 0.81 & 1.00 \\
Analytical  & 0.66 & 0.65 & 0.97 & 0.47 & 0.65 \\
Agreeable   & 0.91 & 0.81 & 0.91 & 0.81 & 0.03 \\
Strategic   & 0.90 & 0.86 & 0.86 & 0.92 & 0.65 \\
\bottomrule
\end{tabular}
\caption{Normalized AgenticPay compatibility matrix used as
held-out ground truth for personality evaluation of LLM dreaming and Greedy.}
\label{tab:personality_gt}
\end{table*}

\begin{table*}[h]
\centering
\small
\begin{tabular}{lr|cc|cc|rr}
\toprule
Task & Requester & \multicolumn{2}{c|}{TKC} & \multicolumn{2}{c|}{Dream Score} & \multicolumn{2}{c}{$\Delta$} \\
     &           & Greedy & Dreaming & Greedy & Dreaming & TKC & Dream \\
\midrule
T01 & Cooperative   & 0.632 & 0.702 & 0.624 & 0.733 & $+0.071$ & $+0.109$ \\
T04 & Competitive   & 0.924 & 0.988 & 0.590 & 0.635 & $+0.064$ & $+0.045$ \\
T08 & Analytical   & 0.654 & 0.695 & 0.595 & 0.592 & $+0.042$ & $-0.003$ \\
T09 & Agreeable    & 0.879 & 0.583 & 0.704 & 0.735 & $-0.296$ & $+0.031$ \\
T17 & Strategic  & 0.890 & 0.802 & 0.598 & 0.641 & $-0.088$ & $+0.043$ \\
\bottomrule
\end{tabular}
\caption{Breakdown for one representative task per requester personality
(5 representatives from the full 25-task set), plus the mean over all
25 tasks. TKC is the held-out ground truth; Dream Score is the LLM
judge's self-rating.}
\label{tab:per-personality-comparison}
\end{table*}

\begin{table*}[b]
\centering
\small
\setlength{\tabcolsep}{6pt}
\begin{tabular}{@{}lp{4.7cm}@{}}
\toprule
\textbf{Setting} & \textbf{Value} \\
\midrule
Task suite & 20 tasks \\
Candidate pool & 20 candidates for RQ1/RQ2; 3 candidates for RQ4 \\
Online-selection budget & 25 rounds \\
Dreaming budget & 30 rounds \\
Consensus criterion & \(K=3\) consecutive selections \\
Evaluation weights & \(\theta=[0.4,-0.1]\) \\
Capability observations & Noise-free oracle observations unless otherwise stated \\
Uniform prior uncertainty & \(\sigma_{\mathrm{init}}=0.25\) \\
LLM backend & GPT-4o \\
Faithfulness protocol & Deterministic MapScore; exploration disabled; 5 seeds\\
\midrule
\textbf{AgenticPay hyperparameters }& \\
\midrule
Discount factor $\gamma$ & $0.99$ \\
Deal success reward $D$ & $30.0$ \\
Deal quality reward $W$ & $55.0$ \\
Round efficiency reward $E$ & $15.0$ \\
\bottomrule
\end{tabular}
\caption{Reproducibility settings used in the evaluation.}
\label{tab:hyperparameters}
\end{table*}

\begin{figure*}[h]
    \centering
    \includegraphics[width=\textwidth]{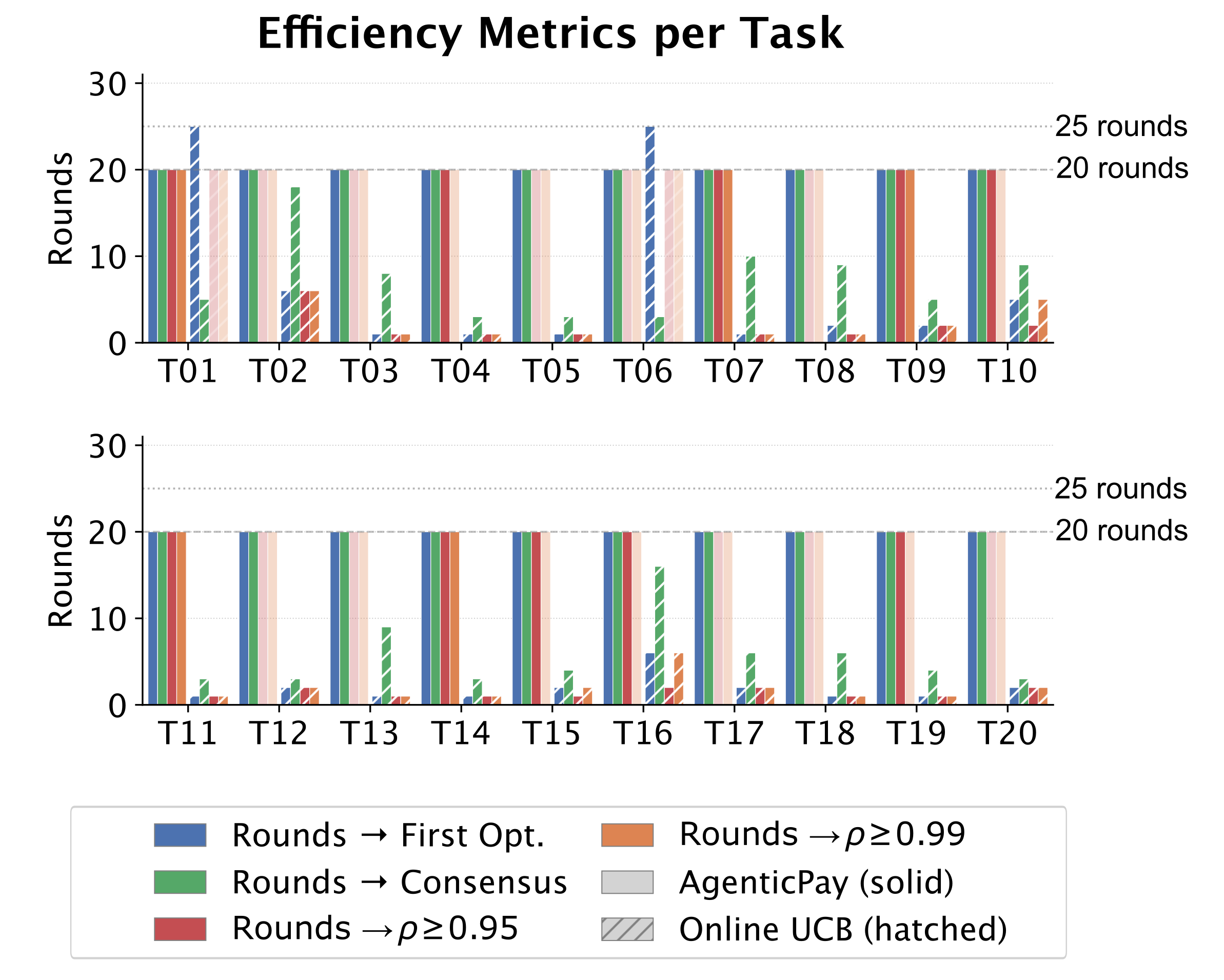}
    \caption{Compare \model's matching efficiency with AgenticPay}
    \label{fig:benchmark_efficiency}
\end{figure*}

\begin{figure*}[h]
    \centering
    \includegraphics[width=\textwidth]{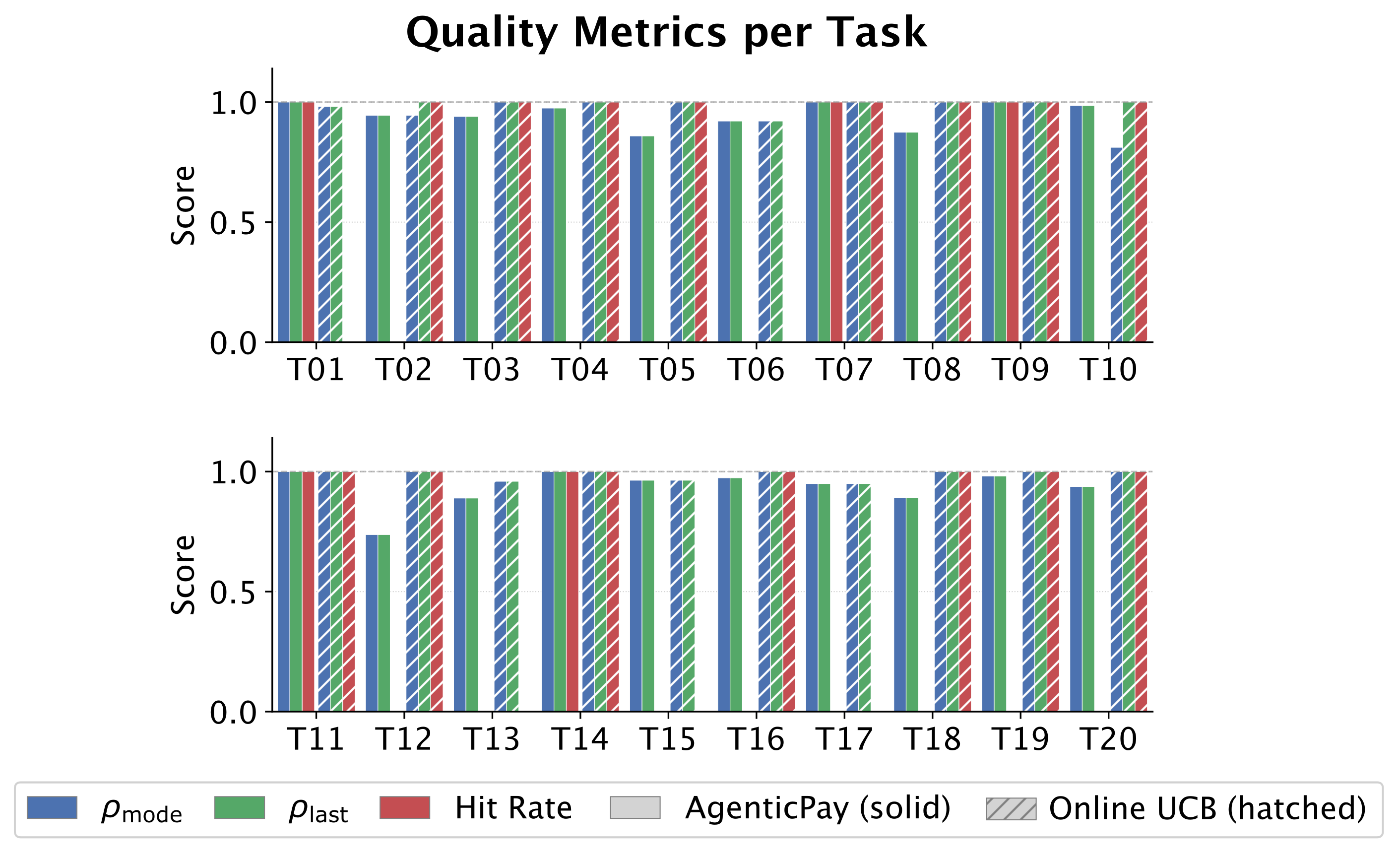}
    \caption{Compare \model's matching quality with AgenticPay}
    \label{fig:benchmark_quality}
\end{figure*}

\begin{figure*}
    \centering
    \includegraphics[width=0.85\linewidth]{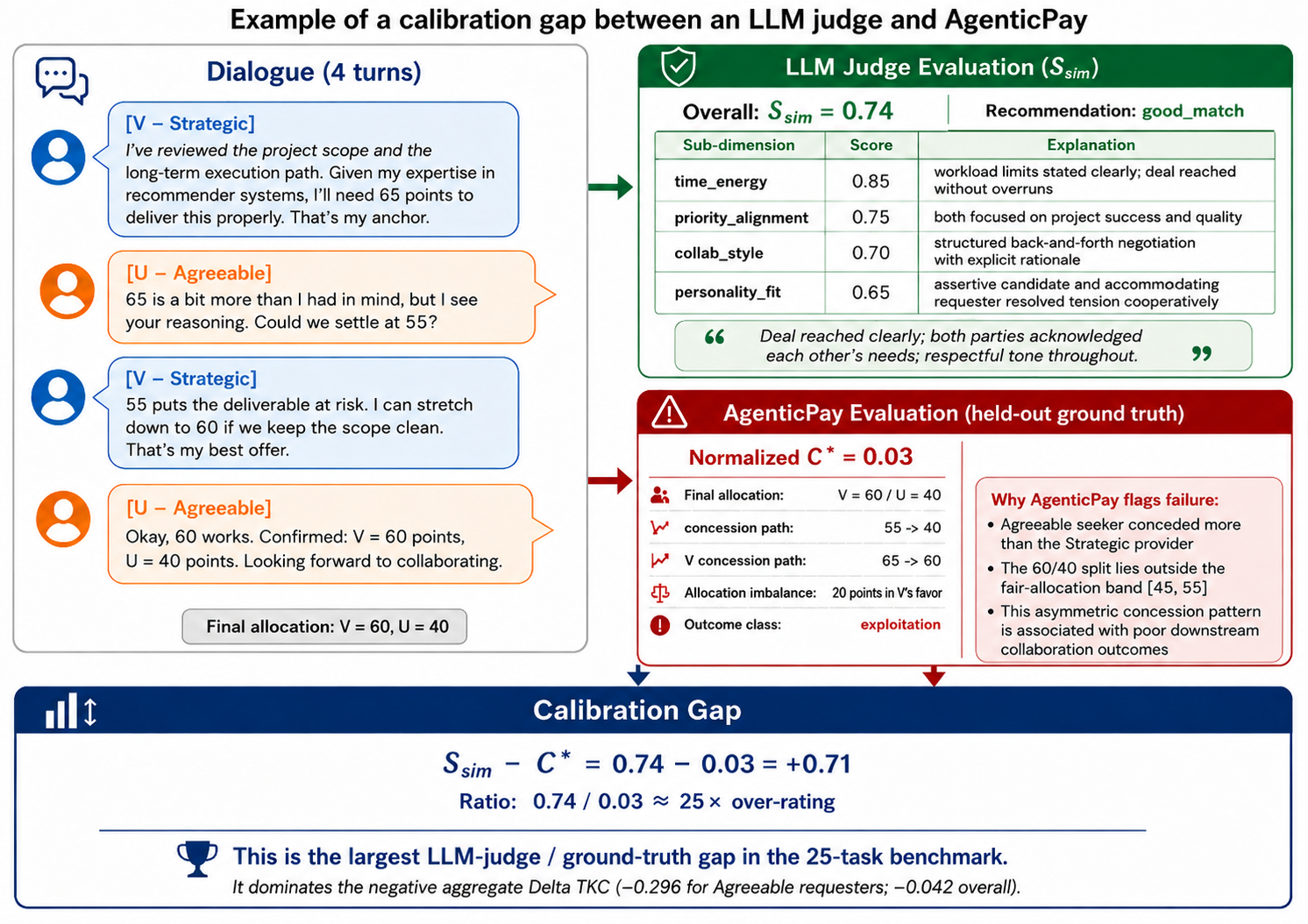}
    \caption{The Agreeable--Strategic trap. The dialogue 
features no impasses, mutual acknowledgment, and explicit agreement on terms---surface features that the LLM judge reads as compatibility ($S_{\text{sim}} = 0.74$, \texttt{good\_match}). AgenticPay's outcome-grounded GlobalScore, however, evaluates the fairness of the final allocation: U receives only 40 points while V receives 60 points, classifying the deal as exploitation of the Agreeable seeker. The Agreeable$\times$Strategic cell scores normalized $C^* = 0.03$. The 0.71 gap is structural: the LLM judge has no mechanism to differentiate between ``reached agreement'' and ``reached \emph{fair} agreement''.}
    \label{fig:agreeable_trap}
\end{figure*}

\clearpage
\section{User Survey Motivation Evidence}
\label{app:user-survey}

This appendix summarizes the HuMAS academic collaboration-agent survey (\(n=16\)) used to motivate the collaboration-matching problem studied in the main paper.
The survey is not used as an evaluation benchmark for \model{}.
Instead, it provides lightweight evidence for three assumptions behind our problem formulation: the target users include early-career researchers with limited collaboration networks, useful collaborators are often complementary rather than similar, and researchers want proactive help in finding people or resources when bottlenecks arise.

\begin{figure}[t]
    \centering
    \begin{minipage}[t]{0.46\linewidth}
        \centering
        \includegraphics[width=\linewidth]{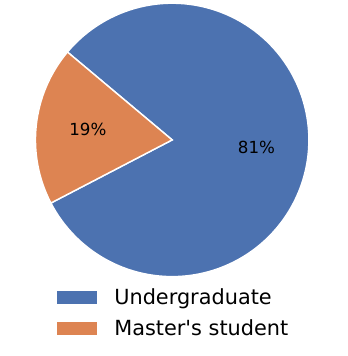}\\[-0.3em]
        {\small (a) Academic role}
    \end{minipage}
    \hfill
    \begin{minipage}[t]{0.49\linewidth}
        \centering
        \includegraphics[width=\linewidth]{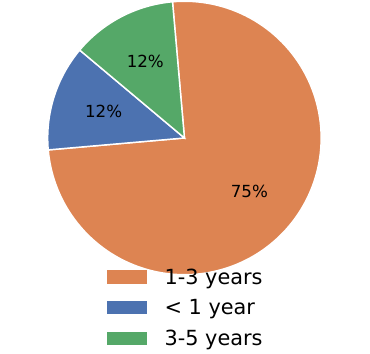}\\[-0.3em]
        {\small (b) Years in current role}
    \end{minipage}
    \caption{Survey respondent profile. (a) Academic role and (b) years in current role. The sample is dominated by early-career researchers: 13 of 16 respondents are undergraduates, and 12 of 16 report 1--3 years in their current role.}
    \label{fig:survey_respondent_profile}
\end{figure}

Figure~\ref{fig:survey_respondent_profile} characterizes the respondent population.
The sample is concentrated among early-career researchers: 13 of 16 respondents are undergraduates, and 12 of 16 report 1--3 years in their current role.
This profile is important for the motivation because early-career researchers often have weaker informal networks for discovering collaborators, mentors, or domain experts.
For this population, a collaboration agent should not only answer questions after a user names a helper; it should help surface plausible collaborators when the user can only describe a task, bottleneck, or missing capability.

\begin{figure}[t]
    \centering
    \begin{minipage}[t]{0.4\linewidth}
        \centering
        \includegraphics[width=\linewidth]{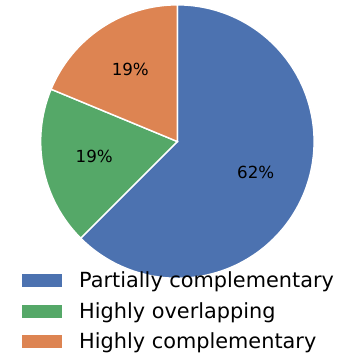}\\[-0.3em]
        {\small (a) Expertise relationship}
    \end{minipage}
    \hfill
    \begin{minipage}[t]{0.55\linewidth}
        \centering
        \includegraphics[width=\linewidth]{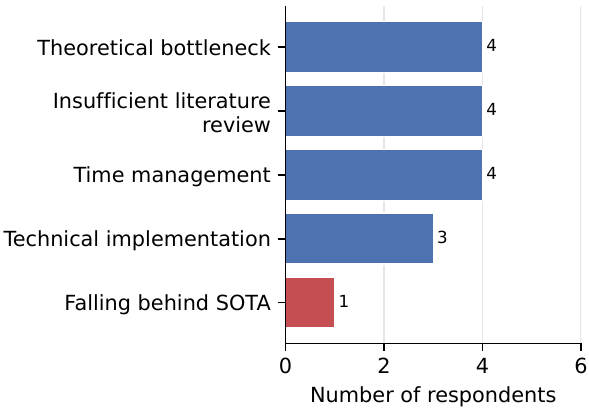}\\[-0.3em]
        {\small (b) Main research challenge}
    \end{minipage}
    \caption{Collaboration patterns and research challenges. (a) Respondents describe their key collaborators as largely complementary: 13 of 16 report a partially or highly complementary expertise relationship. (b) Reported research challenges span theory, literature review, time management, implementation, and tracking state of the art.}
    \label{fig:survey_collaboration_patterns}
\end{figure}

Figure~\ref{fig:survey_collaboration_patterns} supports our use of capability complementarity as a central matching signal.
When asked about their most important collaborator, 13 of 16 respondents described the expertise relationship as partially or highly complementary, while only 3 described it as highly overlapping.
The same respondents reported heterogeneous research challenges, including theoretical bottlenecks, insufficient literature review, time management, technical implementation, and falling behind the state of the art.
These responses suggest that a useful matching system should connect a task to actors who cover the specific missing capability, rather than simply retrieve researchers or agents with similar profiles.

\begin{figure}[t]
    \centering
    \begin{minipage}[t]{0.4\linewidth}
        \centering
        \includegraphics[width=\linewidth]{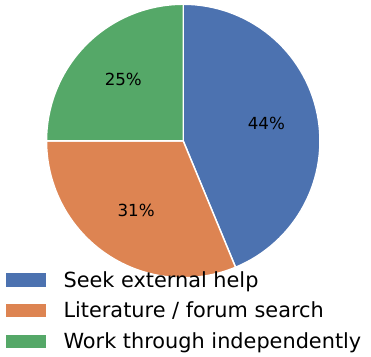}\\[-0.3em]
        {\small (a) First reaction to bottleneck}
    \end{minipage}
    \hfill
    \begin{minipage}[t]{0.55\linewidth}
        \centering
        \includegraphics[width=\linewidth]{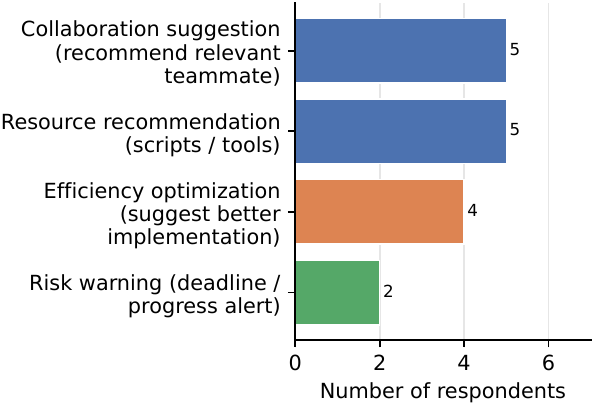}\\[-0.3em]
        {\small (b) Desired AI assistance}
    \end{minipage}
    \caption{Bottleneck responses and desired AI assistance. (a) Seven of 16 respondents report seeking external help from a mentor or expert as their first reaction to a bottleneck. (b) Collaboration suggestions and resource recommendations are the two most frequent desired forms of proactive AI assistance.}
    \label{fig:survey_bottleneck_ai_assistance}
\end{figure}

Figure~\ref{fig:survey_bottleneck_ai_assistance} further motivates proactive collaboration support.
When encountering a bottleneck, 7 of 16 respondents reported that their first reaction is to seek external help from a mentor or expert.
For desired AI assistance, collaboration suggestions and resource recommendations were the two most frequent choices, with 5 respondents selecting each.
This pattern motivates the upstream matching role of \model{}: before downstream coordination or task execution can happen, the system must decide which human, proxy agent, service agent, or reusable skill is worth connecting to the requester.
Because the survey is small and skewed toward early-career respondents, we treat it as motivation rather than generalizable user-study evidence.

\end{document}